\documentclass[runningheads]{llncs}
\usepackage{amsmath}

\usepackage{wrapfig, booktabs}
\usepackage{graphicx}
\usepackage{subfigure}
\usepackage{amsmath,amssymb}
\usepackage{times}
\usepackage{epsfig}
\usepackage{graphicx}
\usepackage{amsmath}
\usepackage{amssymb}
\usepackage{dirtytalk}
\usepackage{xcolor}
\usepackage{cite}
\usepackage{graphicx}
\usepackage{subfigure}
\usepackage{amsmath,amssymb}

\begin{document}
\title{Improving Pneumonia Localization via Cross-Attention on Medical Images and Reports}
\titlerunning{Cross-Attention Pneumonia Localization}
%
\author{Riddhish Bhalodia \inst{1}  \and Ali Hatamizadeh\inst{2} \and Leo Tam\inst{2} \and Ziyue Xu\inst{2} \and Xiaosong Wang\inst{2} \and Evrim Turkbey\inst{3} \and Daguang Xu\inst{2}}

     
%
\institute{School of Computing, University of Utah, UT, USA \\ \email{riddhishb@sci.utah.edu} \and NVIDIA Corporation, USA \\ \email{\{ahatamizadeh, leot, ziyuex, xiaosongw, daguangx\}@nvidia.com}\and Department of Radiology and Imaging Sciences, National Institutes of Health Clinical Center,Bethesda,USA \\ \email{evrim.turkbey@nih.gov}}

\maketitle              
\begin{abstract}
 Localization and characterization of diseases like pneumonia are primary steps in a clinical pipeline, facilitating detailed clinical diagnosis and subsequent treatment planning. Additionally, such location annotated datasets can provide a pathway for deep learning models to be used for downstream tasks. However, acquiring quality annotations is expensive on human resources and usually requires domain expertise. On the other hand, medical reports contain a plethora of information both about pnuemonia characteristics and its location. In this paper, we propose a novel weakly-supervised attention-driven deep learning model that leverages encoded information in medical reports during training to facilitate better localization. Our model also performs classification of attributes that are associated to pneumonia and extracted from medical reports for supervision. Both the classification and localization are trained in conjunction and  once trained, the model can be utilized for both the localization and characterization of pneumonia using only the input image. In this paper, we explore and analyze the model using chest X-ray datasets and demonstrate qualitatively and quantitatively that the introduction of textual information improves pneumonia localization. We showcase quantitative results on two datasets, MIMIC-CXR and Chest X-ray-8, and we also showcase severity characterization on COVID-19 dataset.
\end{abstract}
\section{Introduction}
\label{sec:intro}




Pneumonia localization in chest X-rays and its subsequent characterization is of vital importance. At times pneumonia is a symptom of a disease (e.g., for COVID-19), and the severity and location of pneumonia can affect the treatment pipeline. Additionally, as manual annotations of the disease are resource-heavy, automatic localization can provide annotations or a guide to annotations for downstream deep learning-based tasks that rely on bounding-box annotations. However, to train an automatic pneumonia location detector, we need a large amount of annotated data that is not readily available and presents us with a chicken and egg problem.



Medical reports, on the other hand, are highly descriptive and provide a plethora of information. For instance, a report snippet ``diffused opacity in left-lung" indicates the structure and location of opacity that can indicate pneumonia. Ideally, we can distill such information from the text and inform the localization of corresponding images without added supervision.
The problem of associating textual information and leveraging it to improve/inform image domain tasks (e.g., object detection) is termed visual grounding. In computer vision literature, visual grounding methods have successfully demonstrated such a weak supervision approach of finding accurate annotation with text captions \cite{datta2019align2ground, xiao2017weakly}. Such methods usually rely on a two-stage approach where the first stage is a box detector, and methods like Faster-RCNN \cite{ren2015faster} provide great box-detectors for natural images. However, the luxury of having a good initial box-detector is not available in medical imaging models.

Textual information from medical reports is hard to distill, as they have lot of specific terminologies, no clear sentence construction and many redundant/extraneous information. Therefore, in this paper, we establish a set of attributes that are informative about the location as well as characterization for pneumonia, and automatically extract them. We propose an attention-based image-text matching architecture to train a localization network, that utilizes the extracted attributes and the corresponding images. Additionally, we also have a jointly-trained attribute classification module. Attribute classification provides a more richer information as compared to a simple pneumonia classification score and the bounding box localization. The probability value for each attribute can be used for characterizing new images that might not have clinical reports associated. We showcase quantitative results on two chest X-ray datasets, MIMIC-CXR \cite{johnson2019mimic}, and Chest X-ray-8 \cite{wang2017chestx}. Additionally, we showcase qualitative results on the COVID-19 dataset \cite{cohen2020covid} and demonstrate severity characterization by utilizing the predicted attributes. Although this method is extensively evaluated for pneumonia localization, it can easily be applicable to other diseases as well.

\section{Related Works}
\label{sec:related}

Weakly supervised visual grounding methods utilize textual information from natural image captions and perform object localization. Most such methods \cite{datta2019align2ground, xiao2017weakly} are two-stage methods, where the first stage is a box-detector such as Faster-RCNN \cite{ren2015faster} trained on natural images. Other methods such as Retinanet \cite{lin2017focal} are end-to-end approaches for object localization that rely on pre-defined anchors. Weakly-supervised visual grounding is commonly achieved by enforcing similarity between image and text modalities in native space \cite{chen2018knowledge}, or in a learned common latent space \cite{datta2019align2ground}. Some other works utilize attention models for vision-language alignments \cite{liu2019adaptive}, as well as contrastive loss \cite{gupta2020contrastive}.

 Class activation mapping (CAM) \cite{zhou2016learning} and its variants \cite{selvaraju2017grad} rely on a surrogate classification task to localize the regions of interest (ROI) and have been utilized in the context of disease localization \cite{wang2017chestx}. More recent works incorporate image manipulation to localize the objects better \cite{singh2017hide, choe2019attention}. Another class of methods utilize attention-based mechanisms to match the image and text modalities. Stacked cross-attention (SCAN) \cite{lee2018stacked} is one such approach that provides a heat map of the image, based on how bounding boxes attend to the associated captions. Some recent works \cite{wei2020multimodality} incorporate similar ideas for image-text matching and improve upon SCAN. 
 
Retinanet \cite{lin2017focal} and its variants have seen success in supervised pneumonia localization in Kaggle RSNA Pneumonia detection challenge \cite{kagglersna}. Other works employ semi-supervision for disease localization and use limited amount of annotations \cite{li2018thoracic}. Some recent works \cite{moradi2018bimodal, wu2020automatic} use entire reports and match image and text features for disease classification and utilize CAM model for localization. The proposed method computes ROI-text similarity and weight ROIs for localization, and the attribute classification module can characterize different properties of a given X-ray image.

\section{Methods}
\label{sec:methods}

\subsection{Attribute Extraction}
\label{sec:attr_extract}

For text features we utilize the clinical reports from the MIMIC-CXR dataset \cite{johnson2019mimic}. These reports are involved, complicated, and without clear sentence structures to be utilized directly. Instead, we pre-process this data to extract a dictionary of important text-attributes indicative of pneumonia location and characteristics. These attributes are keywords associated with the disease word in the clinical report. For example, the attribute set for an image with pneumonia can be \emph{left, diffuse, basal}, etc, where each word describes the location, structure or characteristic of pneumonia. Figure \ref{fig:mimic-fig}(green boxes and attributes) show  examples of pneumonia annotation and the associated attributes. We emphasize here that the attribute-set is a set of 22 keywords that is kept constant, this is described in full detail in the supplementary material. We utilize a modified version of natural language-based label extractor provided by \cite{wang2017chestx} to extract these attributes associated with each image and each disease class. A Word2Vec \cite{mikolov2013efficient} model is trained on the entire set of medical reports. Using this trained model we extract the text-features for the limited attribute set. For a given text report T we extract M attributes, and their features are given as $\{\mathbf{m}_i\}_{i=1}^M$.

\subsection{Box detector and Image Features}
\label{sec:data-process}

An initial box-detector is essential for two-stage grounding methods \cite{datta2019align2ground}. Kaggle RSNA Pneumonia detection challenge \cite{kagglersna} provides annotated X-ray images that can be utilized to train a box-detector. It contains ChestX-Ray-8 \cite{wang2017chestx} images, with 2560 pneumonia annotations (bounding -boxes). Retinanet \cite{lin2017focal} and its variations are the best performing architectures in this challenge, therefore, we train a Retinanet with Resnet-50 backbone. This network produces regions of interest (ROIs) and pneumonia classification score. Additionally, we also get the ROI-features coming from the last convolution map of the network. For a given image I, we extract N ROIs (anchor boxes), the features are given as $\{\mathbf{r}_i\}_{i=1}^N$, the classification scores as $\{{s}_i\}_{i=1}^N$, and the geometric information (four box corner coordinates) as $\{\mathbf{g}_i\}_{i=1}^N$. This Retinanet box detector also acts as our \emph{supervised baseline} for evaluating the textual information's contribution to disease localization. Once trained, we fix the weights of the box-detector.


\subsection{Network Architecture and Attention Model}
\label{sec:architecture}

The proposed architecture is shown in Figure \ref{fig:architecture}, the architecture has three major components described as follows:
\begin{figure}[!t]
    \centering
    \includegraphics[width=\textwidth]{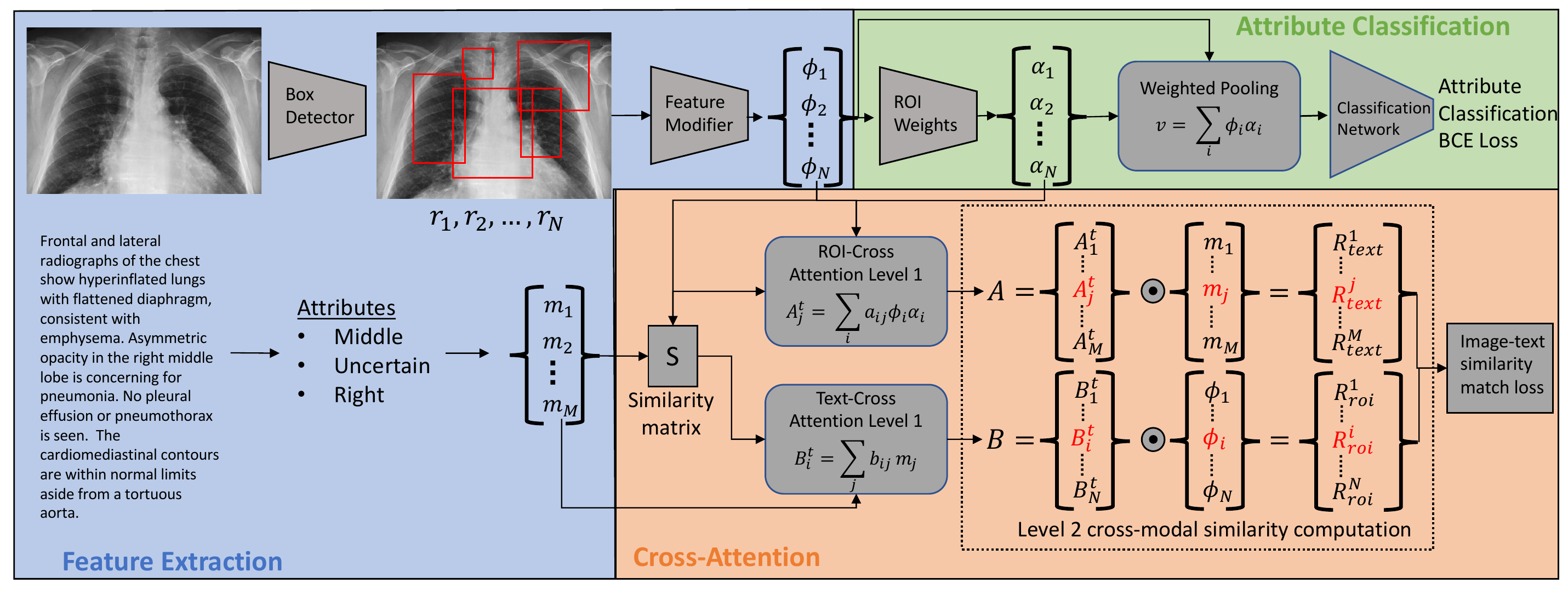}
    \caption{Network architecture for training the attention based image-text matching for localization.}
    \label{fig:architecture}
\end{figure}

\textbf{Feature Extractor}
First the features are extracted for the text and ROI as described in sections \ref{sec:attr_extract} and \ref{sec:data-process}, respectively. As these text and ROI features are coming from fixed architectures and not affected by training they are essentially independent and we need an agency to modify one of them to \emph{match} to the other. For this, we modify the ROI features as, $\boldsymbol{\phi}_i = W_{1}\mathbf{r}_i + W_2[\texttt{LN}(W_g \mathbf{g}_i)| \texttt{LN}(W_s s_i)]$. The $W$'s are the weights applied via fully connected layers which are to be optimized, $\texttt{LN}$ denotes the layer normalization, and $[\cdots|\cdots]$ denotes concatenation.  The modified ROI-features ($\boldsymbol{\phi}_i$) and the text-features $\mathbf{m}_i$ (directly from the Word2Vec model) are of the same dimension. This module is given by blue region in Figure \ref{fig:architecture}.

\textbf{Attribute Classification}
The method's primary goal is to provide selectivity in the ROIs, i.e., the box-detector will extract a large number of ROIs (via the anchors of Retinanet \cite{lin2017focal}) and we select the best ones. We want to discover the appropriate ROIs that can successfully classify the attribute string; for this, we have a two-stage approach (shown in green in Figure \ref{fig:architecture}). First, we compute a weight vector from the ROI-features $\boldsymbol{\phi}_i$, called $\alpha_i$. These provide weights on each ROI feature, and we get an aggregate ROI-feature:  $\mathbf{v} = \sum\limits_{i = 1}^N\alpha_i\boldsymbol{\phi}_i$. Secondly, we utilize $\mathbf{v}$ to perform a multi-label attribute classification by passing $\mathbf{v}$ through a set of fully-connected layers to produce an attribute probability vector. The classification loss is given by binary cross-entropy between the target attributes and the predicted attribute probability vector, we denote it as $\mathcal{L}_{BCE}$. The ROI weights have another function; it also allows for the selection of ROIs that matches with the input attributes best in the cross-attention module described below.

\begin{figure}[!t]
    \centering
    \includegraphics[width=\linewidth]{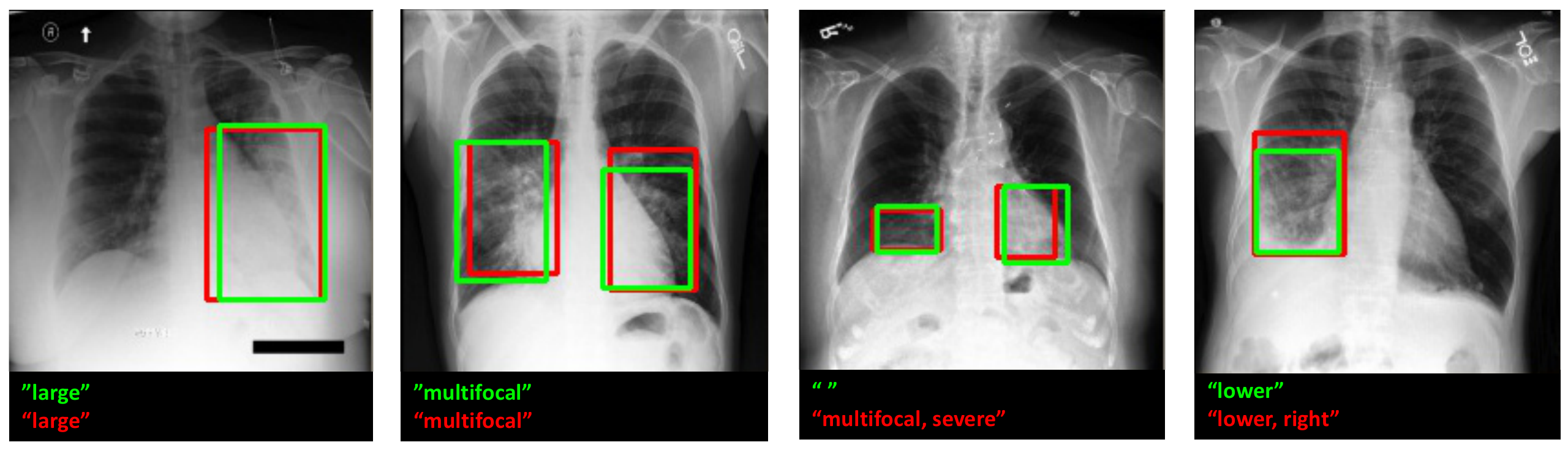}
    \caption{Examples of localization and attribute classification from MIMIC-CXR test data. {\color{green} Green} : expert annotated boxes and extracted attributes, {\color{red} Red }: predicted boxes and attributes.}
    \label{fig:mimic-fig}
\end{figure}
\textbf{Cross-Attention}
The cross-attention is employed to answer the following questions: (i) How well does each weighted ROI describe the input set of text attributes? And (ii) How well does each attribute describe the set of ROIs?. A similar technique is employed in the stacked cross-attention (SCAN) \cite{lee2018stacked}. First, we construct weighted contribution vectors for a given image, attribute pair as, 
    $\mathbf{A}_j = \sum\limits_{i=1}^N \alpha_i\boldsymbol{\phi}_i a_{ij} \quad \text{and} \quad \mathbf{B}_i = \sum\limits_{j=1}^M \mathbf{m}_j b_{ij}$.
Where, $i$ and $j$ denote the index of ROI and attribute respectively. Finally, the $a_{ij}$ and $b_{ij}$ are given by
    $a_{ij} = \frac{\exp{(\lambda_a s_{ij})}}{\sum\limits_j \exp{(\lambda_a s_{ij})}} \quad,\quad b_{ij} = \frac{\exp{(\lambda_b s_{ij})}}{\sum\limits_i \exp{(\lambda_b s_{ij})}}$. The
$\lambda$s are constants, and $s_{ij}$ represents the cosine-similarity between $\boldsymbol{\phi}_i$ and $\mathbf{m}_j$, and is normalized as given in \cite{lee2018stacked}. $\mathbf{A}_j$ represents the aggregate ROI feature based on its contribution to the text attribute $\mathbf{m}_j$, and $\mathbf{B}_i$ represents the aggregate attribute feature based on its contribution to the ROI feature $\boldsymbol{\phi}_i$. Now, we match the features from two modalities, i.e., the weighted ROI feature ($\mathbf{A}_j$) should represent the text attribute ($\mathbf{m}_j$) information (and vice versa). To enforce this another level of similarity is introduced via cosine similarity, i.e., 
$R_{text}^j = \frac{\mathbf{A}_j^T \mathbf{m}_j}{||\mathbf{A}_j|| ||\mathbf{m}_j||} \quad,\quad R_{roi}^i = \frac{\mathbf{B}_i^T \boldsymbol{\phi}_i}{||\mathbf{B}_i|| ||\boldsymbol{\phi}_i||}$.
Mean similarity values are: $S_{roi}(I, T) = \frac{1}{N}\sum\limits_{i=1}^N R_{roi}^i$ and $S_{text}(I, T) = \frac{1}{M}\sum\limits_{j=1}^M R_{text}^j$. These mean similarity values reflect how well a given image I matches with the associated report T.

\subsection{Loss Construction and Inference}
\label{sec:loss}
Along with the classification loss function, we need another loss which enforces the ROI features to be matched with the input attribute features. We propose to use a contrastive formulation by defining \emph{negative} ROIs ($I_n$) and attributes ($T_n$). We use the negative samples and the positive samples ($I, T$) to define similarity ($S_{roi}$ and  $S_{text}$ from previous section) and formulate a triplet loss as follows:

\begin{align}
    \mathcal{L}_{trip} = \max(\beta - S_{roi}(I, T) + S_{roi}(I_n, T), 0) +  \nonumber \\
    \max(\beta - S_{text}(I, T) + S_{text}(I, T_n), 0)
    \label{eq:triplet}
\end{align}
$\beta$ is the triplet loss margin which we set as 0.8 for all experiments. \emph{Negative ROIs} $I_n$ are created by taking the set of lowest ranking ROIs coming from the Retina-net box detector. As Retinanet has pre-defined anchors, we are not filtering out any legitimate boxes when obtaining negative anchors/ROIs. \emph{Negative attributes} $T_n$ are the negatives of individual attributes in the set, i.e., each attribute word will have its corresponding negative attribute word. These are chosen by finding the nearest word to the given attribute using the pre-trained word2vec model. Hence, the final loss is given as $\mathcal{L} = \mathcal{L}_{trip} + \mathcal{L}_{BCE}$.
\textbf{Inference: }We utilize the $\alpha_i$s as weights for ROI selection, and following that we perform non-maximal suppression to remove redundant ROIs. As our inference only depends on $\alpha$ we do not require any text input during the testing.

\begin{table}[!t]
\centering
\caption{ Pneumonia localization performance on different dataset using different methods, the Retinanet \cite{lin2017focal} refers to the supervised baseline. }
\label{tab:loc_mimic}
\setlength{\tabcolsep}{7pt}
\renewcommand{\arraystretch}{1.2}
\begin{tabular}{lllll}
\toprule
\textbf{Method} & \textbf{Dataset} & \textbf{IoU@0.25 } & \textbf{IoU@0.5 } & \textbf{IoU@0.75 }    \\
\midrule
CAM~\cite{zhou2016learning}      & MIMIC-CXR      & 0.521        & 0.212    & 0.015    \\
GradCAM~\cite{selvaraju2017grad}  & MIMIC-CXR      & \textbf{0.545}     & 0.178    & 0.029    \\
Retinanet \cite{lin2017focal}        & MIMIC-CXR             &  0.493       & 0.369 & 0.071           \\
Proposed w/o classification       & MIMIC-CXR             &  0.510       & 0.408 & 0.097           \\
Proposed        & MIMIC-CXR             & 0.529 & \textbf{0.428} & \textbf{0.123}    \\
\hline
Retinanet \cite{lin2017focal}        & Chest X-ray-8             &  0.492       & 0.430 & \textbf{0.115}           \\
Proposed w/o classification        & Chest X-ray-8             &  0.484       & 0.422 & 0.099           \\
Proposed        & Chest X-ray-8             & \textbf{0.507} & \textbf{0.439} & \textbf{0.114}    \\
\bottomrule
\end{tabular}%

\end{table}

\section{Datasets and Implementation Details}
\label{sec:dataset-imp}
The Retinanet trained on the RSNA challenge data (described in section \ref{sec:data-process}) is used as the initial box-detector in all the experiments. For training the proposed model we utilize the MIMIC-CXR dataset \cite{johnson2019mimic}, it is a large-scale dataset that consists of 473,064 chest X-ray images with 206,754 paired radiology reports for 63,478 patients. We process each clinical report associated with a frontal X-ray image to extract the attributes (described in section \ref{sec:attr_extract}). As there are only limited set of clinical reports with the attributes in the fixed set of 22 keywords, we only utilize the images corresponding to pneumonia and having at least one of the attributes in this set, which results in 11,308 training samples. We train the proposed network on this subset of MIMIC images with early stopping at 185 epochs as the validation loss reaches a plateau (less than 0.1\% change in 10 epochs). This trained model is used to perform quantitative and qualitative evaluations in the results section. We divide the data into $90\%, 5\%, 5\%$ as a training, validation and testing split. We would also like to quantify the effect that an attribute classification module would have on the localization performance, therefore, we train another model without the classification module with just the triplet loss (Eq. \ref{eq:triplet}). Additional information about the hyperparameters selection and other implementation details are in supplementary material. For evaluation on MIMIC-CXR we have a held out set of 169 images (part of test split) with pneumonia that are annotated by a board certified radiologist. We also evaluate on Chest-X-ray-8 dataset just utilizing the 120 annotations given for pneumonia as our test set. Finally, we perform a qualitative evaluation on COVID-19 X-Ray dataset \cite{cohen2020covid} that contains 951 X-ray images acquired from different centers across the world and many images have associated medical reports/notes. Due to  different acquisition centers, scanners and protocols across the data the intensity profiles and chest positioning has a huge variation among them.

\section{Results}
\label{sec:results}

\subsection{MIMIC and Chest X-ray-8 Dataset}
\label{sec:mimic}
   We use bounding boxes annotations of MIMIC-CXR images to test pneumonia localization performance (169 annotated images as described in Section \ref{sec:dataset-imp}). For evaluation of localization performance, we employ intersection over union (IoU) evaluated at different thresholds of overlap. The quantitative results are provided in Table \ref{tab:loc_mimic}, where we see that introduction of the textual information improves the IoU score from the supervised baseline. Our method provides a different score for selecting ROIs compared to the supervised baseline, which is trained on a limited and  controlled data that might degrade performance in cases with domain shifts.
   We also see that the proposed network, when trained without the attribute classification module performs worse as compared to one trained with it. Additionally, we also compare against CAM \cite{zhou2016learning} and GradCAM \cite{selvaraju2017grad} that use disease classification to perform localization using the activation heatmap. We threshold the heatmaps to generate the bounding boxes from these methods and use same evaluation as described for proposed method. We see that the proposed method outperforms (especially at 0.5 and 0.75 thresholds indicating better localization) these other weakly-supervised methods. 
   We also use the MIMIC-CXR trained model to evaluate the performance on ChestX-ray-8 \cite{wang2017chestx} dataset, we only use data containing bounding box for pneumonia and not other diseases. The quantitative results using the supervised baseline as well as the proposed method is given in Table \ref{tab:loc_mimic}, and shows the proposed method outperforms the supervised baseline and without classification model. The attribute classification accuracy on the test set for MIMIC-CXR using the proposed method is 95.6\% with an AUC of 0.84. Qualitative localization and attribute prediction is shown in Figure \ref{fig:mimic-fig}.



\begin{figure}[!t]
    \centering
    \includegraphics[width=\linewidth]{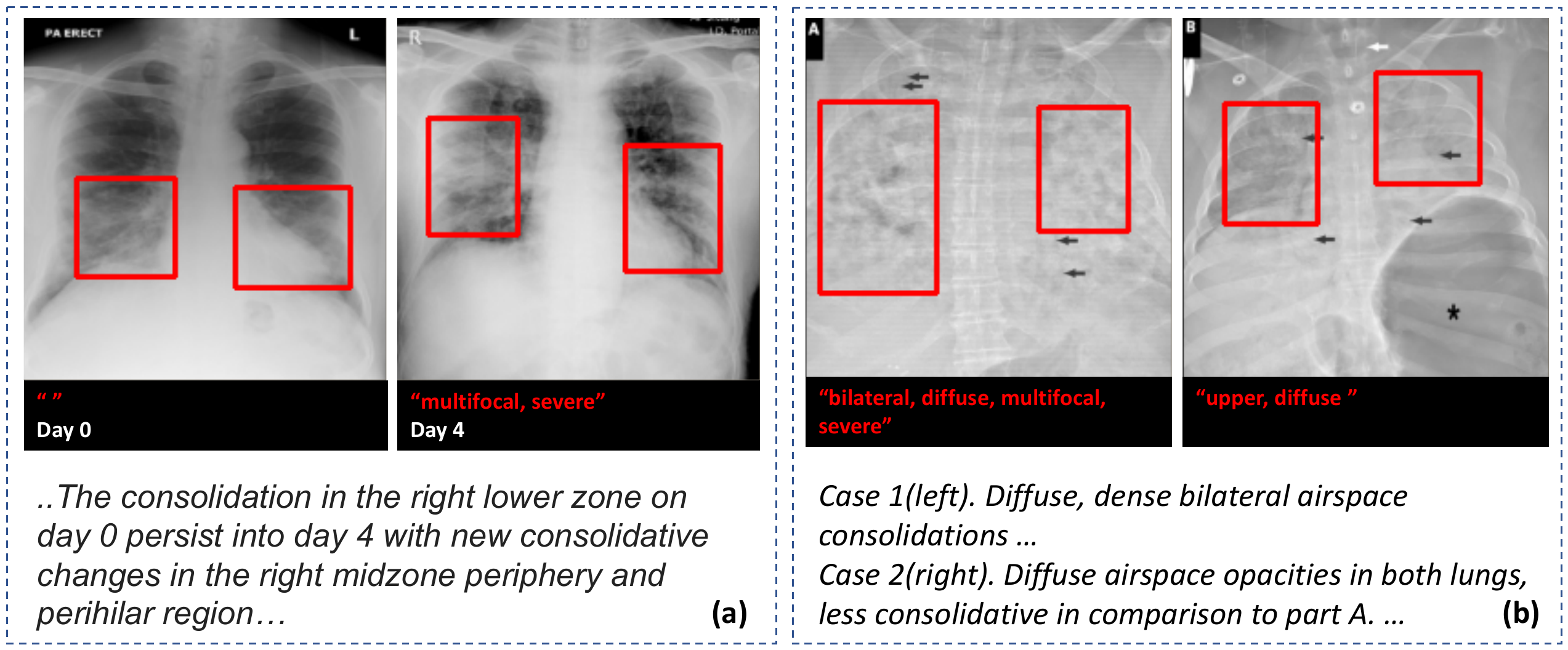}
    \caption{Example case studies for pneumonia characterization from the COVID-19 Chest X-Ray dataset. The images, predicted attributes and localization, report snippet are shown here.}
    \label{fig:covid1}
\end{figure}

\subsection{COVID-19 Dataset}
\label{sec:covid}

\begin{figure}[!t]
    \centering
    \includegraphics[width=\textwidth]{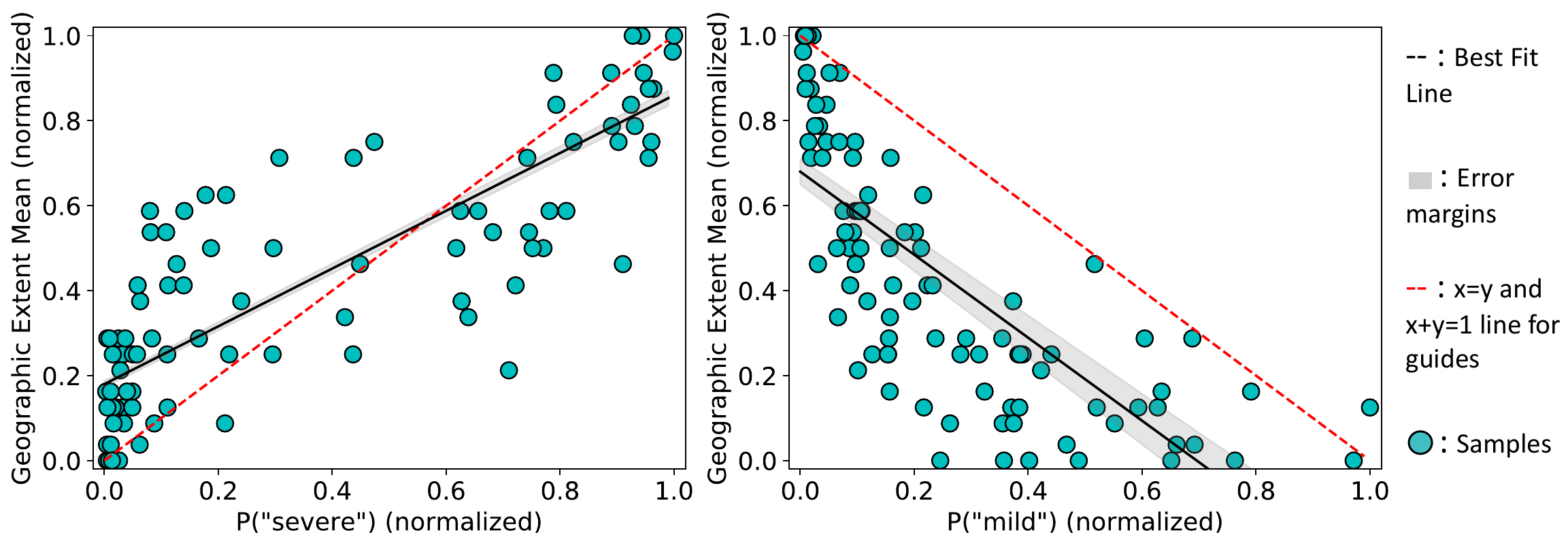}
    \caption{Correlation between the attribute probabilities( P(\emph{mild}) and P(\emph{severe})) and the ground-truth aggregate severity scores from experts.}
    \label{fig:covid2}
\end{figure}
COVID-19 dataset of Chest X-Ray \cite{cohen2020covid} is an important use-case for pneumonia characterization. We use the X-ray images from this dataset as an evaluation set on a model trained on MIMIC data. We look at specific cases and compare them with the reports; for the most part, the predicted attributes align with the information in associated reports. Two such interesting findings are shown in Figure \ref{fig:covid1}. In (a) the two scans of the same subject are shown, these are taken at day 0 and day 4. We notice that our model predicts the attribute (especially ``severe") on day 4 scan, which is also suggestive from the report snippet shown below the images. In (b), despite images being very saturated, our model characterizes well and differentiates two cases as in report snippet.

Another aspect of pivotal importance is being able to characterize the severity of pneumonia. In addition to the Chest X ray images the dataset also provides a measure of severity for a subset of the overall dataset, this subset contains 94 images and their severity is quantified via \emph{geographic extent mean} as described in \cite{cohen2020predicting}. It describes the extent of lung involvement by ground glass opacity or consolidation for each lung. The severity are mean of 3 expert ratings from 0-8 (8 is most severe). We hypothesize that the probability value associated with the attributes ``severe" and ``mild" can describe severity. We compute Pearson and Spearman correlation between the attribute probabilities and the ground truth severity scores, as well as other statistics quantified in Table \ref{tab:covid-stats}. Statistics are evaluated using 5-fold cross validation on 94 cases. The p-values for the computation of each correlation is $< 10^{-10}$ demonstrating a near 100\% confidence that these correlations are statistically significant. This is showcased in Figure \ref{fig:covid2}. A high positive correlation with P(\emph{severe}) and a high negative correlation with P(\emph{mild}) is in line with our hypothesis. 

\begin{table}[!t]
\centering
\caption{Attribute probabilities compared to expert given severity. CC = correlation coefficient, $R^2$ = coefficient of determination, MAE = mean absolute error, MSE= mean squared error}
\label{tab:covid-stats}
\setlength{\tabcolsep}{5pt}
\renewcommand{\arraystretch}{1.2}
\begin{tabular}{llllll}
\toprule
\textbf{Attribute} & \textbf{Pearson CC } & \textbf{Spearman CC } & $\mathbf{R^2}$ &\textbf{MAE} & \textbf{MSE}    \\
\midrule
Severe     & $0.82 \pm 0.001$       & $0.84 \pm 0.005$    &  $0.59 \pm 0.09$   & $0.14 \pm 0.02$    & $0.032 \pm 0.008$ \\
Mild  & $-0.75 \pm 0.003$      & $-0.84 \pm 0.02$        & $0.56 \pm 0.04$    & $0.15 \pm 0.01$    & $0.035 \pm 0.005$ \\
\bottomrule
\end{tabular}%

\end{table}

\section{Conclusion}
\label{sec:conclusion}

This paper introduces a novel attention-based mechanism of leveraging textual information from medical reports to inform disease localization on corresponding images. The network also comprises of a jointly trained attribute classification module. We showcase that the proposed method performs better than the supervised and other weakly supervised baselines. To showcase disease characterization, we test the model on COVID-19 dataset and qualitatively demonstrate that the attributes can characterize the images and perform localization even with extreme image variation. Furthermore, we perform severity characterization using our model that provides a statistically significant correlation with expert given severity ranking of COVID-19 X-rays.

\bibliographystyle{splncs04}
\bibliography{references}

\begin{thebibliography}{10}
\providecommand{\url}[1]{\texttt{#1}}
\providecommand{\urlprefix}{URL }
\providecommand{\doi}[1]{https://doi.org/#1}

\bibitem{chen2018knowledge}
Chen, K., Gao, J., Nevatia, R.: Knowledge aided consistency for weakly
  supervised phrase grounding. In: Proceedings of the IEEE Conference on
  Computer Vision and Pattern Recognition. pp. 4042--4050 (2018)

\bibitem{choe2019attention}
Choe, J., Shim, H.: Attention-based dropout layer for weakly supervised object
  localization. In: Proceedings of the IEEE/CVF Conference on Computer Vision
  and Pattern Recognition. pp. 2219--2228 (2019)

\bibitem{cohen2020predicting}
Cohen, J.P., Dao, L., Roth, K., Morrison, P., Bengio, Y., Abbasi, A.F., Shen,
  B., Mahsa, H.K., Ghassemi, M., Li, H., et~al.: Predicting covid-19 pneumonia
  severity on chest x-ray with deep learning. Cureus  \textbf{12}(7) (2020)

\bibitem{cohen2020covid}
Cohen, J.P., Morrison, P., Dao, L.: Covid-19 image data collection. arXiv
  2003.11597  (2020), \url{https://github.com/ieee8023/covid-chestxray-dataset}

\bibitem{datta2019align2ground}
Datta, S., Sikka, K., Roy, A., Ahuja, K., Parikh, D., Divakaran, A.:
  Align2ground: Weakly supervised phrase grounding guided by image-caption
  alignment. In: Proceedings of the IEEE/CVF International Conference on
  Computer Vision. pp. 2601--2610 (2019)

\bibitem{gupta2020contrastive}
Gupta, T., Vahdat, A., Chechik, G., Yang, X., Kautz, J., Hoiem, D.: Contrastive
  learning for weakly supervised phrase grounding. arXiv preprint
  arXiv:2006.09920  (2020)

\bibitem{johnson2019mimic}
Johnson, A.E., Pollard, T.J., Greenbaum, N.R., Lungren, M.P., Deng, C.y., Peng,
  Y., Lu, Z., Mark, R.G., Berkowitz, S.J., Horng, S.: Mimic-cxr-jpg, a large
  publicly available database of labeled chest radiographs. arXiv preprint
  arXiv:1901.07042  (2019)

\bibitem{lee2018stacked}
Lee, K.H., Chen, X., Hua, G., Hu, H., He, X.: Stacked cross attention for
  image-text matching. In: Proceedings of the European Conference on Computer
  Vision (ECCV). pp. 201--216 (2018)

\bibitem{li2018thoracic}
Li, Z., Wang, C., Han, M., Xue, Y., Wei, W., Li, L.J., Fei-Fei, L.: Thoracic
  disease identification and localization with limited supervision. In:
  Proceedings of the IEEE Conference on Computer Vision and Pattern
  Recognition. pp. 8290--8299 (2018)

\bibitem{lin2017focal}
Lin, T.Y., Goyal, P., Girshick, R., He, K., Doll{\'a}r, P.: Focal loss for
  dense object detection. In: Proceedings of the IEEE international conference
  on computer vision. pp. 2980--2988 (2017)

\bibitem{liu2019adaptive}
Liu, X., Li, L., Wang, S., Zha, Z.J., Meng, D., Huang, Q.: Adaptive
  reconstruction network for weakly supervised referring expression grounding.
  In: Proceedings of the IEEE/CVF International Conference on Computer Vision.
  pp. 2611--2620 (2019)

\bibitem{mikolov2013efficient}
Mikolov, T., Chen, K., Corrado, G., Dean, J.: Efficient estimation of word
  representations in vector space. arXiv preprint arXiv:1301.3781  (2013)

\bibitem{moradi2018bimodal}
Moradi, M., Madani, A., Gur, Y., Guo, Y., Syeda-Mahmood, T.: Bimodal network
  architectures for automatic generation of image annotation from text. In:
  International Conference on Medical Image Computing and Computer-Assisted
  Intervention. pp. 449--456. Springer (2018)

\bibitem{kagglersna}
of~North~America, R.S.: Rsna pneumonia detection challenge (08 2018)

\bibitem{ren2015faster}
Ren, S., He, K., Girshick, R., Sun, J.: Faster r-cnn: Towards real-time object
  detection with region proposal networks. arXiv preprint arXiv:1506.01497
  (2015)

\bibitem{selvaraju2017grad}
Selvaraju, R.R., Cogswell, M., Das, A., Vedantam, R., Parikh, D., Batra, D.:
  Grad-cam: Visual explanations from deep networks via gradient-based
  localization. In: Proceedings of the IEEE international conference on
  computer vision. pp. 618--626 (2017)

\bibitem{singh2017hide}
Singh, K.K., Lee, Y.J.: Hide-and-seek: Forcing a network to be meticulous for
  weakly-supervised object and action localization. In: 2017 IEEE international
  conference on computer vision (ICCV). pp. 3544--3553. IEEE (2017)

\bibitem{wang2017chestx}
Wang, X., Peng, Y., Lu, L., Lu, Z., Bagheri, M., Summers, R.M.: Chestx-ray8:
  Hospital-scale chest x-ray database and benchmarks on weakly-supervised
  classification and localization of common thorax diseases. In: Proceedings of
  the IEEE conference on computer vision and pattern recognition. pp.
  2097--2106 (2017)

\bibitem{wei2020multimodality}
Wei, X., Zhang, T., Li, Y., Zhang, Y., Wu, F.: Multi-modality cross attention
  network for image and sentence matching. In: Proceedings of the IEEE/CVF
  Conference on Computer Vision and Pattern Recognition (CVPR) (2020)

\bibitem{wu2020automatic}
Wu, J., Gur, Y., Karargyris, A., Syed, A.B., Boyko, O., Moradi, M.,
  Syeda-Mahmood, T.: Automatic bounding box annotation of chest x-ray data for
  localization of abnormalities. In: 2020 IEEE 17th International Symposium on
  Biomedical Imaging (ISBI). pp. 799--803. IEEE (2020)

\bibitem{xiao2017weakly}
Xiao, F., Sigal, L., Jae~Lee, Y.: Weakly-supervised visual grounding of phrases
  with linguistic structures. In: Proceedings of the IEEE Conference on
  Computer Vision and Pattern Recognition. pp. 5945--5954 (2017)

\bibitem{zhou2016learning}
Zhou, B., Khosla, A., Lapedriza, A., Oliva, A., Torralba, A.: Learning deep
  features for discriminative localization. In: Proceedings of the IEEE
  conference on computer vision and pattern recognition. pp. 2921--2929 (2016)

\end{thebibliography}

\end{document}


\title{Supplementary: Improving Disease Localization via Cross-Attention on Medical Reports}
%
%
\author{}
%
%
\institute{}
\maketitle              
\section{Attribute Selection}

We utilize a modified version of disease class extractor used in \cite{wang2017chestx} to identify keywords associated with pneumonia on the MIMIC-CXR reports. We plot the histogram of top 100 most frequently occurring keywords/attributes and filter down to the following 21 that describe location, size, structure or important characteristics of pneumonia. All the used attributes are:
\begin{align}
    &\texttt{'left', 'right', 'lower', 'middle', 'upper', 'lateral', } \nonumber \\
        &\texttt{'bilateral', 'basal', 'apical', 'aspiration', 'small', } \nonumber \\
        &\texttt{'large', 'diffuse', 'multifocal', 'focal', 'effusion', } \nonumber \\
        &\texttt{'atelectasis', 'severe', 'acute', 'moderate', 'positive', } \nonumber \\
        &\texttt{'uncertain'} \nonumber
\end{align}

\section{Implementation and Hyperparameter Details}
We utilize Pytorch as our implementation framework, and for training Adam is used for optimization. The loss for the optimization was attribute classification cross-entropy plus the triplet loss (Section 3.4) with equal scaling. All the hyperparameters used are mentioned in Table \ref{tab:param}, the learning rate and weight decay are adopted from \cite{lee2018stacked}. The model is robust to the similarity matrix scales $\lambda$s, and we use cross-validation on triplet loss to compute $\beta$.
Starting at the ROI-features from box-detector, we pass it through a feature modifier layer mentioned in Section 3.3 to obtain $\boldsymbol{\phi}_i$'s. The ROI-weights architecture comprises of two fully connected layers of sizes 1024 and 512 plus an output fully connected layer. The first two layers use Leaky-ReLU as their non-linear activation function and the output layer uses Sigmoid. The output is passed through a softmax function to obtain a probability values ($\alpha_i$'s) over the ROIs. The attribute classification architecture comprises of four fully connected layers of sizes $[512, 512, 256, 128]$ plus an output fully connected layer. The first four layers use Leaky-ReLU followed by batch-normalization and the output layer uses Sigmoid.

\begin{table}[!t]
\centering
\caption{ Values of hyperparameters used for training the proposed model. }
\label{tab:param}
\begin{tabular}{ll}
\toprule
\textbf{Hyperparameter} & \textbf{Value } \\
\midrule
Learning Rate      & $0.0001$        \\
Weight Decay  & $0.0005$     \\
Number of Epochs        & $185$  \\
Batch Size       & $10$      \\
Triplet Loss Scale ($\beta$)       & $0.8$      \\
Similarity matrix factors ($\lambda$s)       & $1$      \\
\bottomrule
\end{tabular}%

\end{table}

\bibliographystyle{splncs04}
\bibliography{references}